\documentclass[aps, prd, preprintnumbers, floatfix, showpacs, superscriptaddress, 10pt]{revtex4}

\usepackage[dvips]{graphics}
\usepackage{amsmath,amsfonts,bm}
\usepackage{epsfig,bm}
\usepackage{graphicx,comment}
\usepackage{dcolumn,color}

\def\beq{\begin{equation}}  
\def\eeq{\end{equation}}  
\def\bea{\begin{eqnarray}}  
\def\eea{\end{eqnarray}}

\newcommand{\be}{\begin{equation}}
\newcommand{\ee}{\end{equation}}
\newcommand{\bear}{\begin{eqnarray}}
\newcommand{\eear}{\end{eqnarray}}
\newcommand{\ba}{\begin{array}}
\newcommand{\ea}{\end{array}}

\begin{document}
\preprint{\today}


\title{Old Quantization,  Angular Momentum, and Nonanalytic Problems}


\author{Nelia~Mann}
\affiliation{Union College, Physics and Astronomy Department, Schenectady, New York 12308, USA}
\author{Jessica~Matli}
\affiliation{Union College, Physics and Astronomy Department, Schenectady, New York 12308, USA}
\author{Tuan~Pham}
\affiliation{University of Wisconsin at Madison, Physics Department, Madison, Wisconsin 53706, USA}
%


\begin{abstract}
We explore the method of old quantization as applied to states with nonzero angular momentum, and show that it leads to qualitatively and quantitatively useful information about systems with spherically symmetric potentials.  We begin by reviewing the traditional application of this model to hydrogen, and discuss the way Einstein-Brillouin-Keller quantization resolves a mismatch between old quantization states and true quantum mechanical states.  We then analyze systems with logarithmic and Yukawa potentials, and compare the results of old quantization to those from solving Schr\"odinger's equation.  We show that the old quantization techniques provide insight into the spread of energy levels associated with a given principal quantum number, as well as giving quantitatively accurate approximations for the energies.  Analyzing systems in this manner involves an educationally valuable synthesis of multiple numerical methods, as well as providing deeper insight into the connections between classical and quantum mechanical physics.
\end{abstract}


\pacs{
}

\maketitle


\section{Introduction}

The origins of quantum mechanics are usually dated to 1905, with the publication of Einstein's work on the photoelectric effect \cite{photoelectric}, even though the study of Schr\"odinger's equation and matrix mechanics---what we usually think of as quantum mechanics---was developed primarily in the mid-1920s \cite{schrodinger, heisenberg}.  During the intervening two decades, quantum mechanics consisted of a series of ad hoc techniques that combined classical reasoning with simple rules for quantization.  Most of us today are familiar with one of the earliest pieces of this story: in 1911 Niels Bohr developed a model of the atom that places electrons in classical circular orbits but quantizes the allowed values of angular momentum, so as to generate a discrete energy spectrum \cite{bohr}.  This model, which accurately produces the energy spectrum of hydrogen despite creating a misleading physical picture, is usually included in introductory modern physics courses.

However, almost no undergraduate courses spend time on the extensions that followed from this initial work.  Instead, they leap awkwardly to a discussion of wavefunctions and Schr\"odinger's equation, without making any real connection between the these ideas and the Bohr model.  Furthermore, these courses often end before reaching later approximation techniques such as the WKB method, which do link the two.  As a result, many students never develop a real understanding of semiclassical approximations, even though such approximations are quite effective at capturing a range of qualitative and quantitative information and still play an active role in serious theoretical physics: consider for example the analyses of strings in curved spacetime backgrounds, where true quantization is not well understood \cite{tseytlin}.  

In the period after Bohr's work was published, Arnold Sommerfeld, William Wilson, and Jun Ishiwara independently developed extensions of the Bohr model \cite{sommerfeld, wilson, ishiwara}.   These models put the electron in a hydrogen atom into classical elliptical orbits, and then imposed two separate quantization conditions on integrals over radial and angular momentum.  This ``old quantization'' model reproduces some aspects of degeneracy in the hydrogen spectrum: multiple different combinations of the two quantum numbers lead to the same energy levels, different energy levels do not have the corrent number of states.  To resolve that difficulty, it is necessary to consider both the Einstein-Brillouin-Keller method \cite{einstein, brillouin, keller, curtis} and the way a quantum mechanical energy spectrum is altered by changing the number of spatial dimensions.  This reveals that the quantum numbers involved in the old quantization analysis should be shifted by half-integers, which this doesn't alter the energy levels of hydrogen but does correct the degeneracy mismatch.  

We present here an analysis of two other systems: confinement in a logarithmic potential and in a Yukawa potential.  These systems were studied in \cite{GMM}, with a focus on states with no angular momentum.  Here, we extend this work to consider non-zero angular momentum, where the lack of degeneracy in the energy spectrum allows us to clearly see how the old quantization states map onto the true quantum mechanical states.  The analysis also shows that some information about the spectra in these systems can be obtained through simple analytic calculations, demonstrating the power of semiclassical analysis. 

For both the logarithmic and Yukawa potentials, finding full the old quantization and Schr\"odinger energy spectra requires a variety of different numerical methods and therefore provides an excellent opportunity to teach students about root finders, numerical integrators, and differential equation solvers.  In fact, the need to synthesize the separate techniques in order to execute both the old quantization and Schr\"odinger's equation calculations creates a more powerful framework for introducing these tools than presenting them separately.  Furthermore, the need to rely on qualitative and analytic analyses to inform the structure of the computations, and the need to understand the numerical uncertainties involved, add depth to the experience.

In Sec. \ref{Hydrogen} of this paper, we review the basics of old quantization and its application to modeling the hydrogen atom.  We also discuss the mismatch between the number of states present at each energy level in old quantization and in an analysis of Schr\"odinger's equation, and how this can be resolved.  In Sec. \ref{Log}, we explore the comparison for a logarithmic potential.  We show that analytically tractable aspects of old quantization give significant qualitative insight into the ``spread'' of different energies associated with the same quantum number, and that a good quantitative agreement can be found when the old quantization conditions are shifted by half-integers.  In Sec. \ref{Yukawa}, we perform similar comparisons for the Yukawa potential.  In addition to supporting the conclusions drawn from studying the logarithmic potential, we explore the existence of a finite number of bound states in this system.  In Sec. \ref{numerics}, we discuss some of the details of the numerical methods used and their pedagogical value, and in Sec. \ref{conclusions}, we summarize our results and suggest possible avenues of future work.

\section{Hydrogen: A State-Counting Problem}
\label{Hydrogen}

We begin by reviewing old quantization in the traditional case of ``hydrogen'' (that is, an attractive Coulombic potential).  We therefore assume we have a particle of mass $m$ in three dimensions, subject to the spherically symmetric attractive potential
\beq
\label{eqn:CoulombPot}
V_{H}(r) = -\frac{C}{r} \, .
\eeq

Classically, angular momentum conservation guarantees that our particle remains confined to a single ``orbital plane,''  which means we can describe the motion using radial and angular coordinates $r(t)$ and $\theta(t)$.  (We can also choose to orient the orbital plane such that the motion is counterclockwise.)  The coordinates $r(t)$ and $\theta(t)$ satisfy the energy and angular momentum conservation equations
\beq
\label{eqn:Econs}
E = \frac{p_r^2}{2m} + \frac{p_\theta^2}{2mr^2} + V(r) = \frac{p_r^2}{2m} + U_{\mathrm{eff}}(r; p_{\theta}) \,
\eeq
and
\beq
\label{eqn:pcons}
p_{\theta} = mr^2\dot{\theta} \, ,
\eeq
where $E$ is the energy, and $p_{\theta}$ is the magnitude of the angular momentum vector.  The object $p_r = m\dot{r}$ is the radial momentum, which isn't conserved, and $U_{\mathrm{eff}}(r; p_{\theta})$ is the ``effective potential,'' which governs the radial motion.  For the Coulomb potential, the bound orbits consistent with these equations are closed ellipses with one focal point at the origin.  

In old quantization (OQ) \cite{sommerfeld, wilson, ishiwara}, we then impose quantization conditions on the action integrals over radial and angular momentum:
\beq
\label{eqn:quant}
hn_r = \oint p_r \, dr, \hspace{.75in} hn_{\theta} = \oint p_{\theta} \, d\theta \, .
\eeq
Here, each integral is performed over a complete orbit, and we consider $n_r, n_{\theta} = 0, 1, 2, 3, \dots$, with $n_r + n_{\theta} > 0$.  Angular momentum conservation reduces the second condition to $p_{\theta} = \hbar n_{\theta}$, and we can use the energy conservation equation to rewrite the first as
\beq
\label{eqn:nrquant}
hn_r = 2\int_{r_{-}}^{r_{+}} \sqrt{2mE - U_{\mathrm{eff}}(r; \hbar n_{\theta})} \, dr \, ,
\eeq
where $r_{\pm}$ are the turning points for the orbit (found by setting $p_r = 0$ in equation \ref{eqn:Econs}).  Performing this integral for the Coulomb potential and solving for energy then gives us
\beq
\label{eqn:OQH}
\mbox{OQ hydrogen spectrum:} \hspace{.5in} E = -\frac{E_R}{(n_r + n_{\theta})^2} ,
\eeq
where $E_R = \frac{mC^2}{2\hbar^2}$ is the Rydberg energy scale.

Two special types of OQ states deserve particular consideration.  If we choose $n_r = 0$, then we are restricting ourselves to circular orbits.  The radial quantization integral disappears, and energy quantization is given by
\beq
\label{eqn:Ecirc}
E =U_{\mathrm{eff}}(r_c; \hbar n_{\theta}), \hspace{.75in} U'(r_c; \hbar n_{\theta}) = 0 \, .
\eeq
Here, $r_c$ is the radius of a circular orbit, which is found by minimizing the effective potential.  (When we use the Coulomb potential in these equations, we recover equation \ref{eqn:OQH}, with $n_r = 0$.)  These are the states used in the ``Bohr model'' of the atom \cite{bohr}, and can also be thought of as the states that maximize angular momentum for a given amount of energy.

On the other hand, if we choose $n_{\theta} = 0$ we are considering ``orbits'' that consist entirely of radial motion.  These are most easily visualized as the limit of a classical orbit in which we allow $p_{\theta} \rightarrow 0$, which leads to the radial quantization integral
\beq
\label{eqn:Erad}
hn_r = 2\int_{0}^{r_{+}} \sqrt{2mE - V(r)} \, dr \, ,
\eeq
with the upper limit of the integral determined by $E = V(r_{+})$.  Again, using the Coulomb potential in this integral leads back to equation \ref{eqn:OQH}, this time with $n_{\theta} = 0$.  Sommerfeld himself considered these states unphysical and discounted them.  However, we know now that the solutions to Schr\"odinger's equation do include states with zero angular momentum, represented by spherically symmetric wavefunctions.

Now consider the radial part of Schr\"odinger's equation, which is
\beq
-\frac{\hbar^2}{2m}\frac{d^2u}{dr^2} + U_{\mathrm{eff}}\left(r; \tilde{p}_{\theta}\right)u(r) = Eu(r) \, , \hspace{.5in} \frac{\tilde{p}_{\theta}^2}{\hbar^2} = \ell(\ell+1) \, ,
\eeq
where $R(r) = \frac{u(r)}{r}$ is the radial wavefunction, and $\ell$ is the quantum number associated with the magnitude of angular momentum (see for example \cite{griffiths, townsend}).  The solutions to this equation consistent with bound state boundary conditions give us the famous result
\beq
\mbox{Schr. hydrogen spectrum:} \hspace{.5in} E = -\frac{E_R}{n^2} \, .
\eeq
Here, $n = 1, 2, 3, \dots$ is the ``principal quantum number,'' and the angular momentum quantum number is constrained to $\ell = 0, 1, 2, \dots, n - 1$.  

Clearly, in this case old quantization leads to the same energy levels as Schr\"odinger's equation, and just as clearly, in old quantization we should identify the principal quantum number as $n = n_r + n_{\theta}$.  However, if we try to compare the number of states at each energy level, we see a discrepancy.  In the Schr\"odinger spectrum, the ground state is unique: we must choose $n = 1$ and $\ell = 0$.  On the other hand, in the OQ argument there are two possibilities: $\{n_r = 0, n_{\theta} = 1\}$ and $\{n_r = 1, n_{\theta} = 0\}$.  This pattern continues beyond the ground state level: at the $n$th level, there are $n$ possible values for $\ell$, but $n+1$ possible values for $n_{\theta}$.  And since the energies of all of these states are the same, it is difficult to see from this perspective which OQ states should be mapped onto which wavefunction solutions to Schr\"odinger's equation.

The solution to this problem lies partially in Einstein-Brillouin-Keller (EBK) quantization \cite{einstein, brillouin, keller, curtis}.  This modifies old quantization by adding a shift in the dependence on the radial quantum number of $\frac{\mu}{4} + \frac{b}{2}$, where $\mu$ is the number of classical turning points in the radial coordinate $r(t)$, and $b$ is the number of hard-wall reflections.  In the case of the Coulomb system (as well as the logarithmic and Yukawa systems we will consider next), we have $\mu = 2$ and $b = 0$.  

Furthermore, consider that in $D$ dimensions the radial wavefunction for a hyperspherically symmetric potential will satisfy the equation
\beq
-\frac{\hbar^2}{2m}\frac{d^2u}{dr^2} + U_{\mathrm{eff}}\left(r; \tilde{p}_{\theta}\right)u(r) = Eu(r) \, , \hspace{.5in} \frac{\tilde{p}_{\theta}^2}{\hbar^2} = \left(\ell + \frac{D}{2} - \frac{3}{2}\right)\left(\ell+ \frac{D}{2} - \frac{1}{2}\right) \, ,
\eeq
where $R(r) = r^{(1-D)/2}u(r)$ is the radial wavefunction.  Given the above form, the quantum mechanical energy spectrum should generally depend on the combination $\ell + \frac{D}{2}$.  

On the other hand, in classical mechanics angular momentum conservation confines motion to a plane, so that the OQ treatment described above is blind to the number of spatial dimensions the system is actually in.  In order to make our OQ results match work in $D \ne 2$ dimensions, we should expect to have to shift the dependence on $n_{\theta}$ by $\frac{D - 2}{2}$.  

Putting these pieces together, in order to get a good mapping between old quantization states and 3-dimensional Schr\"odinger's equation states, we can consider OQ states given by $n_r, n_{\theta} = \frac{1}{2}, \frac{3}{2}, \frac{5}{2}, \dots$, with $n = n_r + n_{\theta}$.  This eliminates \emph{both} the circular orbits \emph{and} the radial motion states, but recovers the correct energy levels and degeneracy structure.

The degeneracy associated with $\ell$ (or $n_{\theta}$)  present in the Coulomb system obscures what is going on with the mapping between the OQ and Schr\"odinger states, and perhaps makes the need for the above argument less than compelling (after all, we recover the correct energy levels from the original OQ argument).  Since this degeneracy is a special feature of the Coulomb potential---a result of the conservation of the Runge-Lenz vector---we should gain additional insight by considering other spherically symmetric potentials.  

\section{The Logarithmic Potential}
\label{Log}

\begin{figure}
\begin{center}
\resizebox{3.3in}{!}{\includegraphics{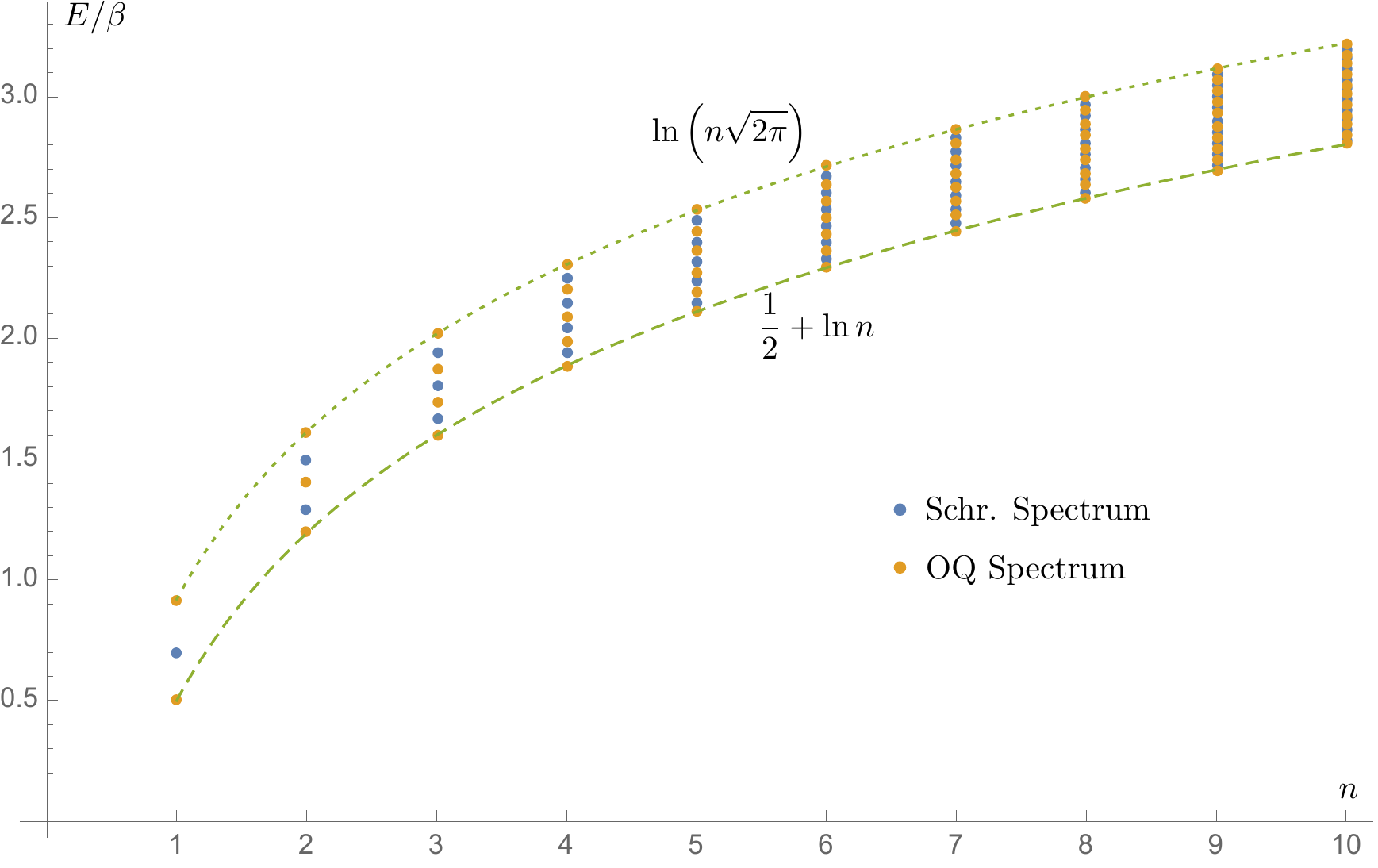}} \hspace{.25in} \resizebox{3.3in}{!}{\includegraphics{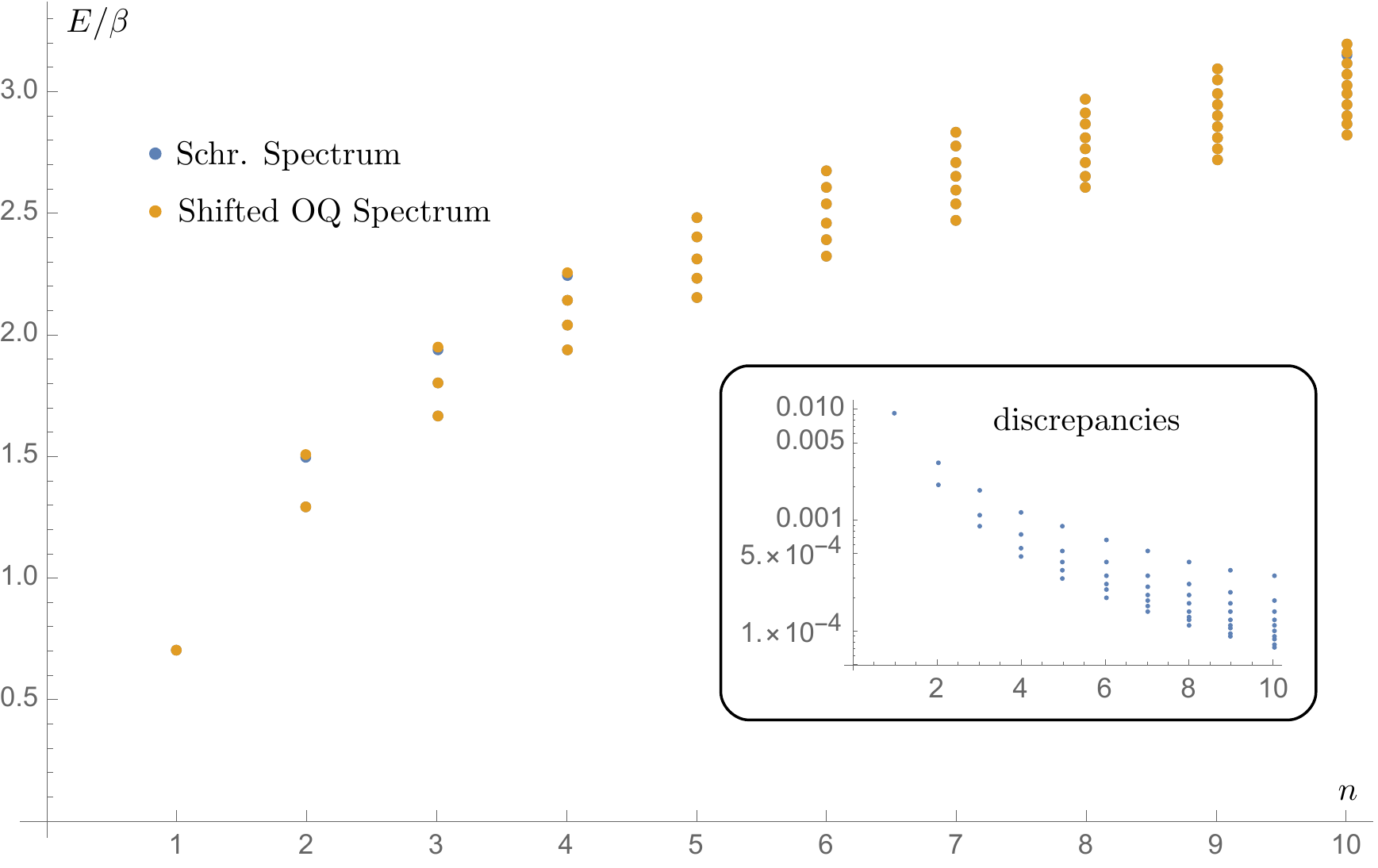}}
\caption{\label{logplot} On the left, a plot showing the old quantization spectrum of the logarithmic potential, together with the Schr\"odinger spectrum.  Lines indicating the parts of the OQ spectrum from radial motion and circular motion are also included.  On the right, a plot showing the shifted old quantization spectrum together with the Schr\"odinger spectrum (the two spectra overlie each other almost exactly).  The inset gives the discrepancies between these values.}
\end{center}
\end{figure}

Suppose we now consider the same particle of mass $m$ in three dimensions, but this time bound in a potential of the form
\beq
V_{\mathrm{log}}(r) = \beta \ln \left(\frac{r}{r_0}\right) \, ,
\eeq
where $\beta$ is a constant that sets the energy scale, and we choose $r_0 = \frac{\hbar}{\sqrt{m\beta}}$ for convenience.  (Shifting the value of $r_0$ simply adjusts the entire energy spectrum upward or downward.)  

Let's begin by using naive old quantization (integer values for $n_r$ and $n_{\theta}$), and start with the circular orbits, specified by $n_r = 0$.  In this case, a straightforward application of equation \ref{eqn:Ecirc} to our logarithmic potential gives an analytic solution:
\beq
\mbox{OQ log Spectrum, $n_r = 0$:} \hspace{.5in} E = \beta\left(\frac{1}{2} + \ln n_{\theta}\right), \hspace{.5in} n_{\theta} = 1, 2, 3, \dots \, .
\eeq

Similarly, we can consider the $n_{\theta} = 0$ scenario.  This problem is also analytically tractable, and following the argument in \cite{GMM}, we can use equation \ref{eqn:Erad} with the logarithmic potential to get
\beq
\mbox{OQ log Spectrum, $n_{\theta} = 0$:} \hspace{.5in} E = \beta\ln \Big(n_{r}\sqrt{2\pi}\Big), \hspace{.5in} n_{r} = 1, 2, 3, \dots \, .
\eeq

Apart from these special cases, finding the spectrum via old quantization requires numerical tools, as does solving Shr\"odinger's equation (see section \ref{numerics}).  Figure \ref{logplot} on the left shows a plot of the results of these analyses, in which we graph the energies vs. the principal quantum numbers.  For each value of $n$, there are multiple different energies, corresponding with different choices of $\ell$ in the exact spectrum and different choices of $n_{\theta}$ in the OQ spectrum.  

We notice immediately that none of the old quantization energies align well with the Schr\"odinger energies: a direct mapping between old quantization states with integer values of $\{n_r, n_{\theta}\}$ and Schr\"odinger states with integer values of $\{n, \ell\}$ is not possible.  However, we also see that for a given value of $n$, the energies of the Schr\"odinger spectrum always lie between the limiting OQ values associated with circular and radial motion.  Furthermore, the larger $n$ gets, the more our Schr\"odinger spectrum seems to ``fill in'' the space between these extremes.  As a result, just by using analytic OQ arguments, we gain significant insight into the ``spread'' of energies associated with one principal quantum number.  Notice in particular that for the logarithmic system this spread is actually a constant, whose value is $\Delta E = \frac{\beta}{2}\left(\ln (2\pi) - 1\right) \approx 0.4189\beta$.

Now consider ``shifted'' OQ states with $n_r, n_{\theta} = \frac{1}{2}, \frac{3}{2}, \frac{5}{2}, \dots$. In this case making the mapping $n_r + n_{\theta} = n$ and $\ell = n_{\theta} - \frac{1}{2}$ creates excellent agreement between the OQ states and the Schr\"odinger states, as is shown in figure \ref{logplot} on the right, and in table \ref{logtable}.  Note that the Schr\"odinger spectrum is consistent with values found previously \cite{logspectra}, and the semiclassical analysis of this system is consistent with that performed in \cite{loganal1}.  Of course, even with the shift modification of the OQ system we do not recover the true quantum mechanical energy spectrum.  The agreeement is strong only for large quantum numbers; for more moderate values additional corrections are necessary \cite{loganal2}.  

\begin{table}
\begin{center}
\begin{tabular}{|c|c|c|c|c|}
\hline
$n$ \ & \ $\ell$ \ & \ $E_{\mathrm{Schr.}}$ \ & \ $E_{\mathrm{OQ}}$ (shifted) \ & \ discrepancies \ \\
\hline
\hline
1 & 0 & 0.697759 & 0.706894 & 0.009135 \\
\hline
2 & 1 & 1.29457 & 1.29659 & 0.00202 \\
 & 0 & 1.50087 & 1.50423 & 0.00336 \\
\hline
3 & 2 & 1.66674 & 1.66759 & 0.000856 \\
 & 1 & 1.80437 & 1.80551 & 0.001133 \\
 & 0 & 1.94304 & 1.94488 & 0.001842 \\
\hline
4 & 3 & 1.93757 & 1.93804 & 0.00047 \\
 & 2 & 2.04086 & 2.04143 & 0.00057 \\
 & 1 & 2.14437 & 2.14511 & 0.000743 \\
 & 0 & 2.24913 & 2.25033 & 0.001199 \\
\hline
5 & 4 & 2.15054 & 2.15084 & 0.000296 \\
 & 3 & 2.23321 & 2.23355 & 0.000344 \\
 & 2 & 2.31592 & 2.31633 & 0.000413 \\
 & 1 & 2.39902 & 2.39956 & 0.000534 \\
 & 0 & 2.48336 & 2.48421 & 0.00086 \\
\hline
6 & 5 & 2.32606 & 2.32626 & 0.000203 \\
 & 4 & 2.39497 & 2.3952 & 0.00023 \\
 & 3 & 2.46387 & 2.46414 & 0.000265 \\
 & 2 & 2.53293 & 2.53324 & 0.000316 \\
 & 1 & 2.60243 & 2.60284 & 0.000407 \\
 & 0 & 2.67308 & 2.67374 & 0.000655 \\
\hline
\end{tabular}
\caption{\label{logtable} A table of the energies derived from Schr\"odinger's equation, as compared with shifted OQ energies for the logarithmic potential, with the identifications $n_{\theta} = \ell + \frac{1}{2}$, and $n_{r} = n - \ell - \frac{1}{2}$.}
\end{center}
\end{table}

\section{The Yukawa Potential}
\label{Yukawa}

\begin{figure}
\begin{center}
\resizebox{4in}{!}{\includegraphics{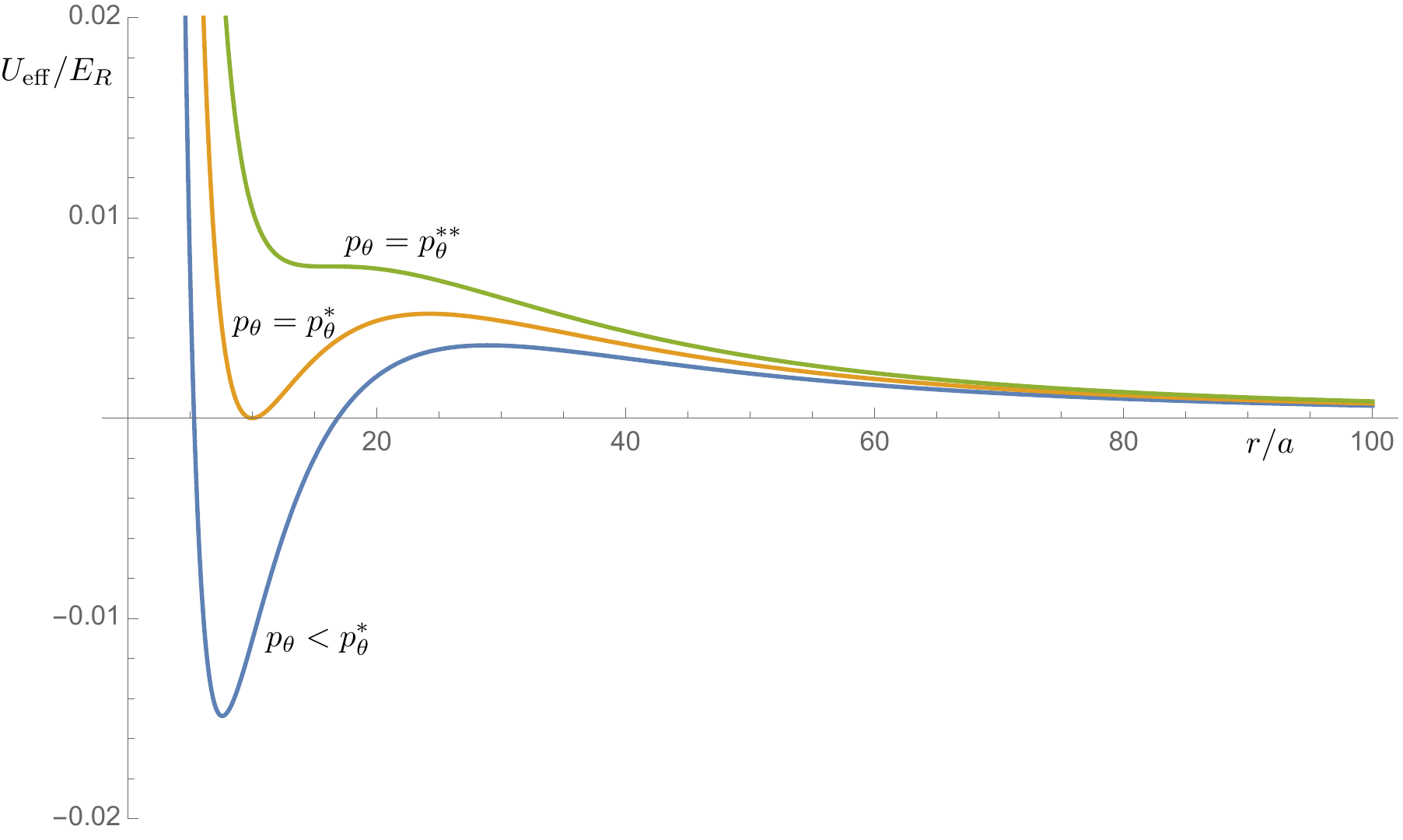}}
\caption{\label{Ueff} The effective potential for $\lambda = 10$, with choices $p_{\theta} < p_{\theta}^{*}$, $p_{\theta} = p_{\theta}^{*}$, and $p_{\theta} = p_{\theta}^{**}$.}
\end{center}
\end{figure}

We can also consider a potential of the form
\beq
V_{\mathrm{Y}}(r) = -\frac{C}{r} \, e^{-r/R} \, ,
\eeq
where $C$ and $R$ are constants (once again with a particle of mass $m$ in three dimensions).  It is convenient to think of this as a modification of the Coulomb potential (equation \ref{eqn:CoulombPot}), in which the exponential suppression factor introduces a rough ``cutoff'' associated with the length scale $R$.  This creates a system in which there will be only a finite number of bound states.  For $r \ll R$, the exponential factor is insignificant, and as a result we expect to have some states that sit close to the origin and are approximately Coulombic.  However, states sitting far away from the origin experience essentially zero force, and should therefore not be bound.  Because the close-in states will have a size controlled by the ``Bohr radius'' $a = \frac{\hbar^2}{mC}$, the number of bound states will be determined by the ratio $\lambda = \frac{R}{a}$.

The inclusion of non-zero angular momentum further complicates this story.  Consider the classical system, focussing on the effective potential shown in figure \ref{Ueff}.  For small values of $p_{\theta}$, the system allows for bound states with negative energies.  However, at a critical value of $p_{\theta}^{*} = \hbar\sqrt{2\lambda} \, e^{-1/2}$, these states disappear and we have $U_{\mathrm{eff}} > 0$ for all $r$.  At values of angular momentum just above this critical point, it is still possible to have classical bound states with positive energies (although tunneling should prevent the existence of quantum mechanical bound states in this regime).  Finally, above a second critical value $p_{\theta}^{**}$ the effective potential ceases to have a minimum at all, so there are no longer any bound states in either the classical or the quantum mechanical system.

Now consider the OQ states corresponding to circular orbits, with $n_r = 0$.  These states are found by minimizing the effective potential (equation \ref{eqn:Ecirc}), which for the Yukawa potential leads to a nonanalytic problem.  However, we can identify a critical value for the quantum number $n_{\theta}$,
\beq
n_{\theta}^{*} = \frac{p_{\theta}^{*}}{\hbar} = \sqrt{2\lambda} \, e^{-1/2} \, .
\eeq
A state with $n_{\theta} > n_{\theta}^{*}$ can no longer have a negative energy, meaning we do not expect it to correspond to a true quantum mechanical bound state.  Similarly, we can look for the old quantization states corresponding to purely radial motion, with $n_{\theta} = 0$.  These are found through equation \ref{eqn:Erad}, which again is not analytically tractable.  However, following the arguments in \cite{GMM}, we can also determine a critical value for the quantum numbers $n_r$:
\beq
n_r^{*} = 2\sqrt{\frac{\lambda}{\pi}} \, .
\eeq
Again, states with $n_r > n_r^{*}$ no longer have negative energy.  

\begin{figure}
\begin{center}
\resizebox{3.3in}{!}{\includegraphics{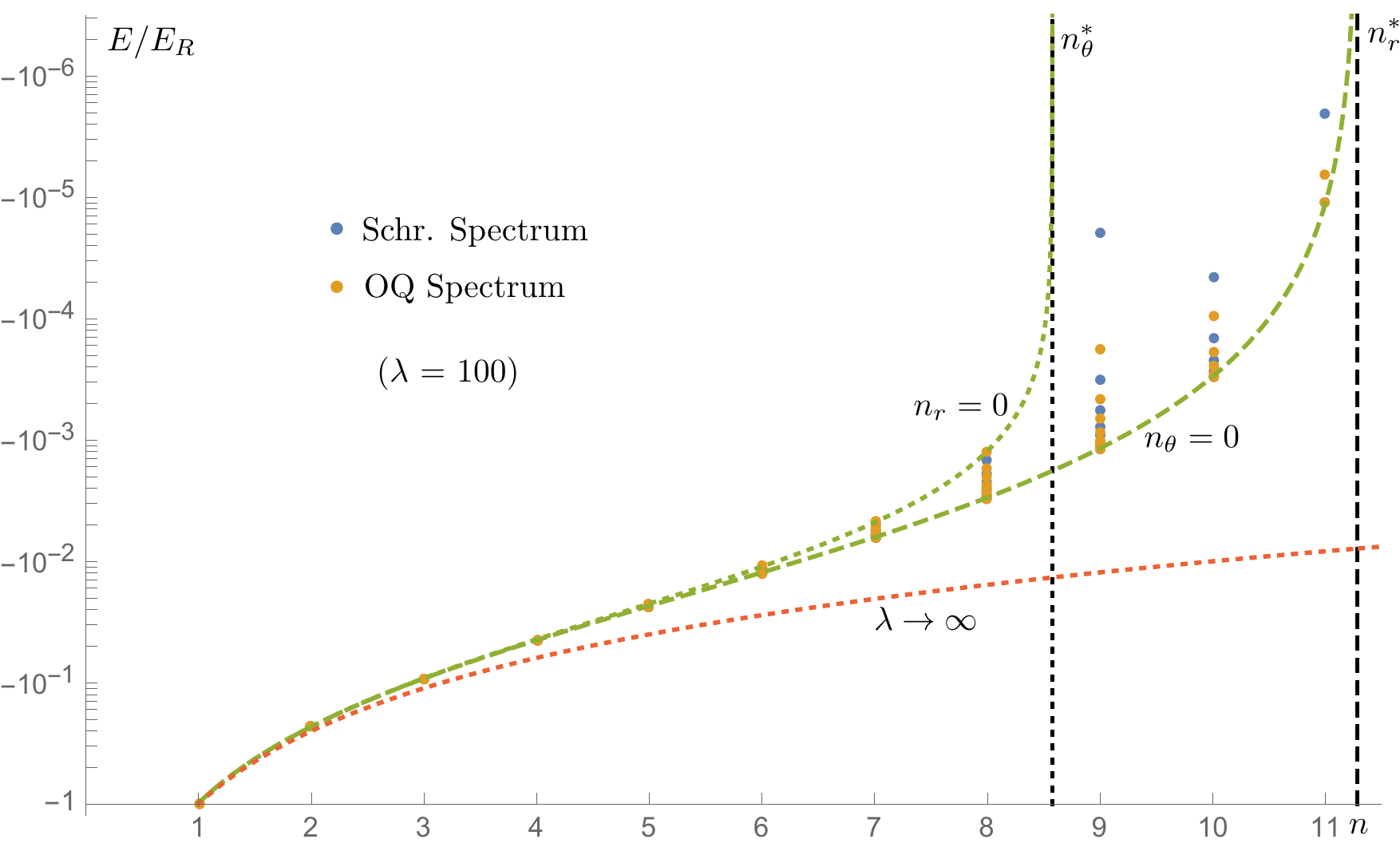}} \hspace{.25in} \resizebox{3.3in}{!}{\includegraphics{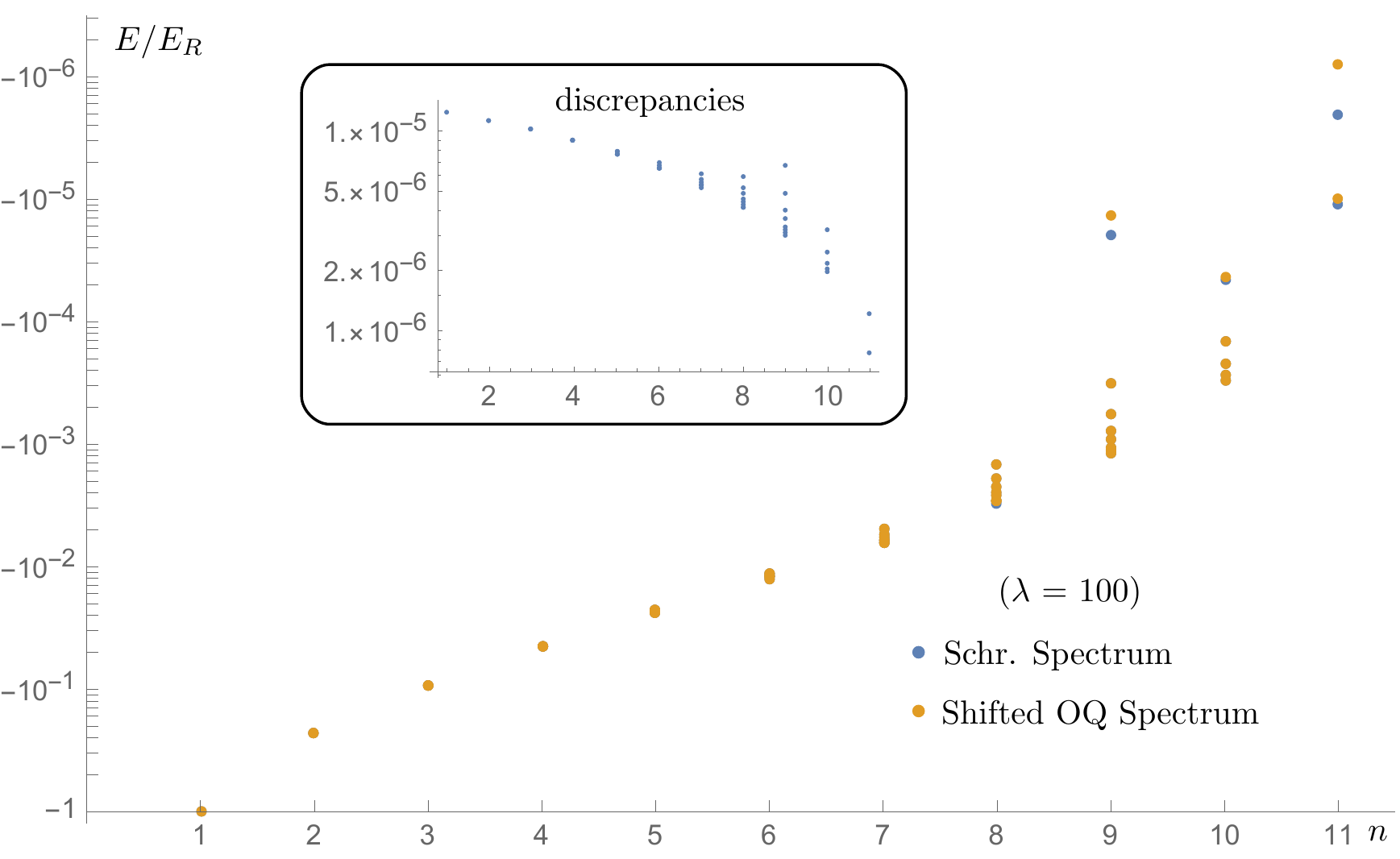}}
\caption{\label{yukplot} On the left, a plot showing the old quantization spectrum of the Yukawa potential, together with the Schr\"odinger spectrum, for $\lambda = 100$.  Additional lines indicate the parts of the OQ spectrum from radial motion and circular motion, the critical values $n_r^{*}$ and $n_{\theta}^{*}$, and the Coulomb spectrum.  On the right, a plot showing the shifted old quantization spectrum together with the Schr\"odinger spectrum.  The inset graph gives the discrepancies.}
\end{center}
\end{figure}

Figure \ref{yukplot} on the left shows the (naive) old quantization spectrum and the exact spectrum, both obtained numerically.  The vertical axis is presented on a logarithmic scale, so as to better show the behaviors of the spectra when the energies are close to zero.  This plot displays many of the same qualitative features discussed for the logarithmic case.  Again, there is no clear mapping between old quantization states corresponding to integer values of $\{n_r, n_\theta\}$, and Schr\"odinger states, and again the Schr\"odinger states always lie between the circular orbit OQ states and the radial motion OQ states.  It is therefore still possible to think of the spread of possible quantized energies corresponding to a given choice of $n$ as limited by these two extremes, though here the lack of analytic solutions makes this insight less powerful.  

On the other hand, the two critical values $n_r^{*}$ and $n_{\theta}^{*}$ do provide us with analytically accessible information that gives us insight into the Schr\"odinger system: we can see that for $n < n_{\theta}^{*}$, there are $n - 1$ bound states associated with principal quantum number $n$.  However, once we are in the region $n_{\theta}^{*} < n < n_{r}^{*}$, there are fewer than $n-1$ bound states for each $n$, and for $n > n_r^{*}$, there are no longer any bound states at all.

Finally, note that a line indicating the hydrogen spectrum (corresponding to $\lambda \rightarrow \infty$) is also included for reference.  For small values of $n$, all of the Yukawa states lie close to the equivalent hydrogen states.  As $n$ becomes larger, the differences become more apparent---in particular that the Coulomb spectrum has an infinite number of negative energy bound states, while the Yukawa spectrum does not.  

We can again obtain a much better agreement between the old quantization states and the Schr\"odinger quantum mechanical states if we shift the OQ quantum numbers, requiring $n_r, n_{\theta} = \frac{1}{2}, \frac{3}{2}, \frac{5}{2}, \dots$, with $n = n_r + n_{\theta}$ and $\ell = n_{\theta} - \frac{1}{2}$.  Figure \ref{yukplot} on the right shows the Schr\"odinger spectrum with the shifted OQ spectrum, and table \ref{yuktable} gives (some of) the numerical data.  The semiclassical spectrum here is consistent with that in \cite{yukvaluesanal}, and the Schr\"odinger's spectrum is consistent with that that found in \cite{lambda2}.  Note that the discrepancies generally become smaller as $n$ increases, but the energies themselves do as well.   Close to the limit where bound states cease to exist, the discrepancies are comparable with the energies, so that the mapping begins to break down.

\begin{table}
\begin{center}
\begin{tabular}{|c|c|c|c|c|}
\hline
$n$ \ & \ $\ell$ \ & \ $E_{\mathrm{Schr.}}$ \ & \ $E_{\mathrm{OQ}}$ (shifted) \ & \ discrepancies \ \\
\hline
\hline
1 & 0 & -0.980149 & -0.980137 & 0.0000122 \\
\hline
2 & 1 & -0.230491 & -0.230479 & 0.0000114 \\
 & 0 & -0.230587 & -0.230575 & 0.0000114 \\
\hline
3 & 2 & -0.0921229 & -0.0921126 & 0.0000103 \\
 & 1 & -0.0923062 & -0.0922959 & 0.0000103 \\
 & 0 & -0.0923977 & -0.0923874 & 0.0000103 \\
\hline
4 & 3 & -0.0441975 & -0.0441885 & $9.1 \times 10^{-6}$ \\
 & 2 & -0.0444556 & -0.0444465 & $9.1 \times 10^{-6}$ \\
 & 1 & -0.0446268 & -0.0446178 & $9.0 \times 10^{-6}$ \\
 & 0 & -0.0447122 & -0.0447032 & $9.0 \times 10^{-6}$ \\
\hline
5 & 4 & -0.0225323 & -0.0225244 & $7.9 \times 10^{-6}$ \\
 & 3 & -0.0228508 & -0.022843 & $7.8 \times 10^{-6}$ \\
 & 2 & -0.0230874 & -0.0230796 & $7.8 \times 10^{-6}$ \\
 & 1 & -0.0232441 & -0.0232363 & $7.7 \times 10^{-6}$ \\
 & 0 & -0.0233221 & -0.0233144 & $7.7 \times 10^{-6}$ \\
\hline
6 & 5 & -0.0112818 & -0.011275 & $6.8 \times 10^{-6}$ \\
 & 4 & -0.0116455 & -0.0116389 & $6.7 \times 10^{-6}$ \\
 & 3 & -0.0119316 & -0.011925 & $6.6 \times 10^{-6}$ \\
 & 2 & -0.0121433 & -0.0121368 & $6.5 \times 10^{-6}$ \\
 & 1 & -0.0122832 & -0.0122768 & $6.5 \times 10^{-6}$ \\
 & 0 & -0.0123528 & -0.0123464 & $6.4 \times 10^{-6}$ \\
\hline
\end{tabular}
\caption{\label{yuktable} A table of some of the Schr\"odinger energies as compared with shifted OQ energies for the Yukawa potential, for $\lambda = 100$, with the identifications $n_{\theta} = \ell + \frac{1}{2}$, and $n_{r} = n - \ell - \frac{1}{2}$.}
\end{center}
\end{table}

\section{An Exploration of Numerical Methods}
\label{numerics}

In the previous two sections, we have argued that results obtained via old quantization can give substantial qualitative insight into the true quantum mechanical spectrum (derived from Schr\"odinger's equation), as well as leading to quantitatively reasonable approximations once we take into account the half-integer shifts.  But a project such as this one can also serve an alternative educational purpose, as a chance for older undergraduate students to explore numerical methods.  Most of the physics underlying this work is drawn from a standard undergraduate curriculum (classical orbits, energy eigenstates in Schr\"odinger's equation, and the Bohr model of the atom), and in the courses a typical upper division student has taken, they will have learned to apply this physics to analytically tractable problems.  However, in order to execute the calculations for more generic potentials they must use a variety of basic numerical tools, including root finders, integrators, and differential equation solvers (see for example \cite{franklin}).  Furthermore, they must assemble these tools in a sophisticated manner in order to obtain the data sets needed to compare OQ and Schr\"odinger energy states.

To perform the calculations involved in the OQ arguments, one must use a root finder to determine the turning points in an orbit, and a numerical integrator to execute the radial quantization integral.  This combination produces a way to find the right-hand side of equation \ref{eqn:nrquant}, given values of energy and angular momentum.  Then, one must again use a root finder to identify values of energy corresponding to quantized choices for $n_r$ and $n_{\theta}$.  We used a basic bisection routine for root finding and the trapezoid method for integration, although there are more sophisticated tools that could be used with students who have more background in numerical methods.  On the other hand, if one wanted to emphasize the way the tools are combined without spending time on the details, one could also utilize pre-written algorithms in programs such as Mathematica or MatLab. 

To obtain the exact energy spectra we employed two different methods and compared the results.  We began by using the ``finite difference method,'' which approximates the Hamiltonian operator as a large, sparse matrix, thus allowing one to use linear algebra algorithms to find eigenvalues (we used prewritten code for the last part).  We also used the ``shooting method,'' in which an initial-value-driven differential equation solver (RK4, in our case)  is combined with a root finder, and the energy eigenvalue is adjusted until the wavefunction satisfies appropriate boundary conditions.  These techniques provide comparable numerical results, but have different pedagogical strengths.  The shooting method allows one to focus on the connection between the spatial localization of a bound state (as expressed in boundary conditions) and the quantization of energy.  On the other hand, the finite difference method creates a valuable link between the ``wavefunction'' approach to quantum mechanics, which students generally learn first, and the ``matrix mechanics'' interpretation often introduced in an advanced quantum mechanics course.  

The most educationally valuable aspect of the numerical work involved in this project was not the individual pieces, but the conceptual mastery involved in synthesizing them: the methods had to be created to feed into each other, not as separate pieces.  In the case of the OQ calculations this process was particularly involved, and beginning by working through the calculation for the Coulomb potential with pen and paper provided a useful guide.  Furthermore, additional effort was required to fully automate the process of finding a large number of energy values (in the case of the Yukawa potential, the code was written to find all bound state solutions for a given choice of $\lambda$), and to adjust numerical parameters to achieve the desired accuracy in the results.  All of these elements are important aspects of applying numerical methods to real research problems, and are often left out of coursework designed to introduce computation.

Finally, applying the numerical algorithms to our systems required a sophisticated blend of qualitative and quantitative reasoning, leading to a deeper understanding of the systems involved.  For example, using a bisection routine generally requires the input of an ``initial bracketing'': a region inside which exactly one root exists.  To use the bisection routine to find $\rho_{\pm}$ for the Yukawa potential, we made use of the fact that these two values would always lie inside the values of $\rho_{\pm}$ for the Coulomb potential, with one value on either side of the location of the minimum.  Similarly, to automate the process of finding all energies associated with a particular principal quantum number, we used the fact that all of the values would lie between the OQ ``radial motion'' and ``circular motion'' cases.

\section{Conclusions}
\label{conclusions}

The old quantization techniques developed by Sommerfeld, Wilson, and Ishiwara, along with the more accurate EBK method, are generally left out of an undergraduate education in quantum mechanics---even though they can often lead to both qualitative and quantitative insights, and in fact are still useful in current research.  In this project, we explored the connection between OQ analysis and Schr\"odinger analysis through the lens of the logarithmic and Yukawa potentials, by using a series of numerical techniques.

We began by reviewing the structure of old quantization in the context of the Coulomb potential.  This allowed us to see that while the traditional OQ argument produces the correct energy levels and is consistent with the existence of degeneracy, it does not clarify how to map OQ states onto Schr\"odinger states precisely.  To solve the problem, it is necessary to consider the more correct EBK method, which shows us that we need to shift the quantum numbers appearing in the OQ analysis by half-integers.  In order to explore this further, we considered the logarithmic and Yukawa potentials.  There, we noted that the ``limiting cases'' of OQ analysis---corresponding to circular and radial motion---effectively bracket the states given by Schr\"odinger's equation.  Thus, OQ analysis can provide substantial insight into the spread of energy states associated with loss of degeneracy; this insight was particularly significant for the logarithmic potential, where the limiting cases are approachable by straightforward analytic calculations.  For the Yukawa potential, analytic tools sufficed only to make predictions about the points where states begin to become unbound.  We also showed that by implementing the half-integer shifts in quantum numbers required by EBK quantization, we could clearly map OQ states onto Schr\"odinger states, so that the OQ results would provide quantitatively accurate approximations.  

Although the ability of OQ analysis to provide quick, pen-and-paper approximations to problems that are otherwise not analytically tractable is often emphasized, in this case both the OQ energy states and the Schr\"odinger states were generally provided by numerical means.  In fact, the range of numerical methods used and the complexity of assembling them makes this type of project ideal as a way to introduce computational physics to older students, while simultaneously giving them greater insight into quantum mechanics.  

It would be interesting to extend this project to consider the states in the Yukawa potential that are classically bound, but with positive energies.  These states exist at the same energy and angular momentum as scattering states, and quantum mechanics allows for tunneling between the two.  In the true quantum mechanical system, these should therefore correspond not to bound states, but instead to resonances, and a mapping might be created between the OQ states and these resonances.  One could also extend this project in a straightforward way by choosing other potentials to explore---for example, a potential in which tunneling would exist between multiple bound states, instead of between a bound state and a scattering state.  Alternatively, one might consider potentials directly connected to physical systems and incorporate comparison with experimental data into the analysis.

\acknowledgements



\end{document}